\documentclass[12pt,a4paper]{article}
\pdfoutput=1
\usepackage[english]{babel}
\usepackage[latin1]{inputenc}
\usepackage[T1]{fontenc}
\usepackage{amsfonts,amsbsy,bm,euscript,mathrsfs}
\usepackage{amssymb,faktor,slashed}
\usepackage{color}
\usepackage[tbtags]{amsmath}
\usepackage[bookmarks=true,colorlinks=true,linkcolor=black,citecolor=black,urlcolor=black,bookmarksnumbered,linktocpage=true]{hyperref}
\usepackage{graphicx}
\usepackage{chngcntr}
\usepackage{makecell}
\usepackage{mathtools}

\usepackage{afterpage}

\usepackage[parsep]{collref}

\usepackage[a4paper,text={170mm,257mm},centering]{geometry}

\numberwithin{equation}{section}

\makeatletter
\renewcommand\section{\@startsection {section}{1}{\z@}%
{-3.5ex \@plus -1ex \@minus -.2ex}%
{2.3ex \@plus.2ex}%
{\normalfont\large\bfseries}}
\renewcommand\subsection{\@startsection{subsection}{2}{\z@}%
{-3.25ex\@plus -1ex \@minus -.2ex}%
{1.5ex \@plus .2ex}%
{\normalfont\normalsize\bfseries}}
\makeatother

\expandafter\def\expandafter\bfseries\expandafter{\bfseries\ifmmode\else\boldmath\fi}
\expandafter\def\expandafter\mdseries\expandafter{\mdseries\ifmmode\else\unboldmath\fi}
\expandafter\def\expandafter\normalfont\expandafter{\normalfont\ifmmode\else\unboldmath\fi}

\providecommand{\href}[2]{#2}

\newcommand{\mathsym}[1]{{}}
\def\id{\protect{{1 \kern-.28em{\rm l}}}}
\def\be{\begin{eqnarray}}
\def\ee{\end{eqnarray}}

\def\ha{\tfrac{1}{2}}
\def\td{\tilde}
\def\ci{\cite}
\def\N{{\mathcal N}}

\def\a{\alpha}

\def\aa{{\a'}}
\def\g{\gamma}

\def\k{\kappa}

\def\K{{\rm S}}

\def\l {\lambda}

\def\const{{\rm const}}

\def\m{\mu}

\def\rS{{\rm S}}

\def\foot{\footnote}
\newcommand{\rf}[1]{(\ref{#1})}

\def\F{{\cal F}}

\def\no{\nonumber}

\def\la{\label}
\def\l{\lambda}

\def\adss{$AdS_5 \times S^5$\ }

\def\p{\phi}

\def\varpi{{\rm w}}

\def\del{\partial}

\def\eps{{\epsilon}}
\def\n{\nu}

\def\ed{\end{document}}

\def\iffa{\iffalse}

\def\te{\textstyle}

\def\dd {{\rm d}}
\def\bb{{\rm b}}

\def\L{\mathcal{L} }\def\pp{{\rm p}}

\def\ov{\over}

\def\k{\varkappa}

\def\bb{{\rm b}}

\def\ka{{\kappa}}

\def\M{{\cal M}}

\def\ed{\end{document}}
\def\F{H}

\def\g{\gamma}

\def \foot{\footnote}\def \ci{cite}\def \l {\lambda}\def \iffa {\iffalse}
 \def \ov {\over }\def \a  {\alpha} \def \ha {{1\ov 2}}
\def \ed {\end{document}}

\def \la {\label}
\def \adss {${\rm AdS}_5 \times S^5~$ }

\def \ov {\over}

\def \ha {{1 \over 2}}
\def \td {\tilde}
\def \ci {\cite}

\def \N  {{\cal N}}
\def \aa  {{\rm a}}
\def \te {\textstyle}

\def \aa {{\rm a}} \def \eps {\epsilon}\def \rr {{\rm r}}

\def \ha {\tfrac{1}{2}} \def  \we {\wedge} 

\begin{document}
\ \hfill{\small Imperial-TP-AT-2022-03 }

\today
\vspace{0.5cm}
\vspace{2.5cm}

\begin{center}

{\Large\bf    On   type  IIB   supergravity  action
 on $M^5 \times X^5$   solutions}

\vspace{1.5cm}
{
S.A.  Kurlyand$^{a,}$\footnote{\ e-mail: kurliand.sa18@physics.msu.ru   } and
A.A. Tseytlin$^{b,}$\footnote{\ Also on leave from  Institute for Theoretical and Mathematical Physics (ITMP)  and Lebedev Institute, \

  \ \ \ e-mail:  tseytlin@imperial.ac.uk}
}

\vspace{0.8cm}
{ \vspace{0.15cm}
$^{a}$ \em Physics Department,  Moscow  State University
\vspace{0.15cm}\\
 \vspace{0.15cm}
$^{b}${\cal{ Blackett Laboratory, Imperial College, London SW7 2AZ, U.K.}}
}

\end{center}

\vspace{1cm}

\begin{abstract}
While the 10d  type IIB    supergravity action evaluated  on AdS$_5\times S^5$  solution  vanishes, the 
5d  effective action  reconstructed  from equations of motion using $M^5 \times S^5$  compactification ansatz 
 is proportional to the   AdS$_5$ volume. The latter  is consistent  with the conformal anomaly interpretation in AdS/CFT   context.  We  show that  this paradox  can be  resolved if,   in the case of  $M^5 \times X^5$  topology,    the 10d  action  contains  an  additional   5-form dependent  ``topological''  term  $\int F_{5M} \wedge F_{5X}$. The presence of this term is suggested also  by gauge-invariance considerations in the PST formulation of  type IIB supergravity action.  We  show  that    this term   contributes to  the 10d action  evaluated on the   D3-brane solution.  
 \end{abstract}
 

\newpage 

\tableofcontents
\setcounter{footnote}{0}
\setcounter{section}{0}
\begin{center}{\large \bf   } \end{center}


\def \ads {AdS$_5$\ }  \def \rr {{\rm r}}  \def \aa  {{\rm a}}\def \L {\Lambda}
\def \OO {{\cal O}}\def \vol {{\rm vol}}

\def \ed { \bibliographystyle{JHEP} \bibliography{Boundary.bib} \end{document}  }
 \def \top  {{\rm top}}
\def \vk {\varkappa}
\def \tf {\tfrac} 
\def \rw {{\rm w}}
 \def \rM  {{\rm M}}  \def \rq  {{\ q}} \def \rS {{\rm S}}
 \def \aa  {{\rm a}}  \def \bb {{\rm b}}
\def \PST {{_{\rm PST}}}    \def \adstr  {${\rm AdS}_3 \times S^3\times  T^4 $  \ } 
\def \K  {X}
\def \rv {{\rm vol}} 
 \def \V {{\rm V}}

\def \aa  {a} \def \bb {b}
\def \M {{\widetilde L }}
\def \F {{\cal F}}

\def \vol {\, {\rm vol}}
\def \k {\kappa} 

\section{Introduction}

Many discussions  of   applications of the maximally   supersymmetric  case   of AdS/CFT duality  \ci{Maldacena:2003nj}
start with a  classical action of 5d   gauged supergravity or simply 5d  gravity   with  a cosmological term 
\be \la{1}
S_5 = - {1 \ov 2 \ka_5^2} \int  d^5 x \sqrt{g} \,  \Big( { R}_5   + 12 L^{-2}     + ... \Big)  \ . 
\ee
Evaluating this action on the \ads  vacuum  solution  with radius $L$ gives a factor of volume of \ads 
space.   Assuming  $S^4$ as a boundary of  AdS$_5$,   
 the   regularized  value of the  volume 
 reproduces the planar part  of the  UV divergent (conformal  a-anomaly)
   term in   the  free energy of $\N=4$   $SU(N)$   super Yang-Mills theory  on $S^4$  (see, e.g.,\ci{Liu:1998bu,Henningson:1998gx,Gubser:1998vd,Blau:1999vz,Burgess:1999vb,Russo:2012ay,Giombi:2020kvo})\foot{\la{f1}
   Here
    we use that 
   ${1\ov 2 \ka_5^2} = {L^4 {\rm vol} (S^5) \ov 2 \ka_{10}^2}$,
   $ 2 \ka_{10}^2  = { (2 \pi)^7 g_s^2 \a'^4}$, $L^4 = 4 \pi g_s  \a'^2 N $.
   To recall, ${\rm vol}(S^n)= { 2 \pi^{n+1\ov 2} \ov \Gamma({n+1\ov 2})}\to_{_{n=5}} =   \pi^3$, 
      ${\rm vol}({\rm AdS}_{2n+1})= { 2(-1)^n  \pi^{n} \ov \Gamma( n+1)} \log (\Lambda  r) \to_{_{2n+1=5}} =   \pi^2\log (\Lambda  r) $  and 
   $R_5= -20 L^{-2}$. 
     $\rr$  is the radius of  boundary 4-sphere    and 
   $\L$ is  an  IR cutoff on the AdS side (corresponding to   UV cutoff on the  SYM side). 
    }
\be  S_5=  {8L^4  \ov 2 \ka_5^2} {\rm vol}({\rm AdS}_{5}) =     N^2  \log ( \L \rr)  \ . 
\la{0} \ee
The action like \rf{1}   is  also  a  starting point of investigations  of  AdS black hole thermodynamics  \ci{Hawking:1982dh,Witten:1998zw, Chamblin:1999tk,Emparan:1999pm,Chamblin:1999hg}.

The  5d  gauged supergravity action    is    assumed  to follow    from the 10d   type IIB supergravity action compactified   on $S^5$  \ci{Gunaydin:1984qu,Pernici:1985ju,Kim:1985ez}. However, 
  the actual  compactification 
procedure involves  starting with the   10d  field equations \ci{Schwarz:1983qr,Howe:1983sra}, 
substituting there an $S^5$ compactification ansatz   and then reconstructing  the corresponding action for the 5d fields
 (cf.  \ci{Cvetic:1999xp,Lu:1999bw,Cvetic:2000nc,Baguet:2015sma}).  
The  bosonic part of the  10d type  IIB action  may be written as\foot{Here    $|F_p|^2 = {1 \ov p!} F_{\mu_1 ...\mu_p} F^{\mu_1 ...\mu_p}$. Extra $\ha $ in the normalization of the $F_5$ kinetic term 
 has to do with the requirement that  the corresponding analog of the Einstein equation 
   should contain  the contribution of the   stress tensor of only   the  self-dual half  of $F_5$.}
\begin{align}
\hat S_{10}=-&\frac{1}{2\kappa_{10}^2}\Big\{\int\dd^{10}x\sqrt{G}\Big( e^{-2\phi}\Big[R+4(\partial_\mu\p)^2-\tfrac{1}{2}{|H_3|}^2\Big] \no \\
&\ \ \ \ \ \qquad\qquad \qquad\qquad -\ha  |{{F}_1}|^2   -\ha  |{{ F}_3}|^2 -  \tfrac{1}{ 4}  |{{F}_5}|^2  \Big) -\ha \int B_2 \we  F_3 \we F_5 \Big\} + ...\ , \la{2}\\
& F_1 = d C_0 \ , \qquad  F_3 = dC_2  - C_0   H_3 \ , \qquad 
  F_5= d C_4 - \ha C_2 \we  H_3   +   \ha B_2 \we  F_3 \ . \la{3} 
\end{align}
Here, as usual,   the  self-duality condition  $F_5 = {}^*F_5$  is    relaxed \ci{Bergshoeff:1995sq}
 and   
is   imposed   by hand 
at  the level of equations of motion (alternative approaches   that involve 
    auxiliary fields  where the self-duality condition   follows  from  the equations of  motion are discussed in 
\ci{Henneaux:1988gg,Schwarz:1993vs,Pasti:1996vs,DallAgata:1997gnw,DallAgata:1998ahf,Belov:2006jd,Sen:2015nph,Mkrtchyan:2019opf,Mkrtchyan:2022xrm}).

Comparing   \rf{1} and \rf{2}     we arrive at the following apparent paradox:  the 10d action \rf{2} evaluated on the vacuum \adss   solution
\be ds_{10}^2 = L^2 \big(ds^2_{\rm AdS_5}  + d \Omega^2_5\big)
\ , \ \ \ \ \ \ \ \ \    
 F_5 = 4 L^{-1}   ( \eps_5 + {}^* \eps_5)\  \la{4}, \qquad  L^4= 4 \pi \a'^2 g_s N  \ ,   \ee 
  is clearly vanishing 
($R= - 20 L^{-2} + 20 L^{-2}=0, \  |F_5|^2=0$)\foot{Note that a self-dual 5-form is real  in the case of Minkowski 10d signature
but  is imaginary  in the Euclidean signature case.}
  while  the  value of the  5d action \rf{1}   on  the \ads    solution is  non-zero    \rf{0}   and consistent    with  the  AdS/CFT  duality.

  It is of course well known that substituting  some   special-symmetry  ansatz for a subset of  fields 
     into the action is not the same as  
  doing  this in  the equations of motion  and then reconstructing the corresponding  dimensionally reduced  action
  for the remaining  field variables. However, the   values of the actions  on the   full solutions  are expected to match. 
    Furthermore, the  problem  is that  the 10d  action 
   and, in particular,  its on-shell  value  should be 
     more fundamental:   it should  follow  from (a properly defined)  quantum string theory path integral. 
  Thus  using the   10d approach  is 
    important if one is  to go beyond the leading  order in $\a'$, in  particular,   in the context of AdS/CFT.

  One may wonder if this  issue  has  to do  with  the subtlety 
   of implementing  self-duality  of $F_5$. 
  However, this is not the case: 
 similar  disagreement between  the on-shell values of the  reduced   3d  action  and the 10d  action
  is found  in the case of ${\rm AdS}_3 \times S^3\times T^4 $  background   supported   by a 3-form flux.
  Here the 10d action is   well defined off-shell  for a generic  3-form field and the 
    effective 6d self-duality of the  latter (implying the vanishing of the 10d action)  is just a feature of a particular solution.
  
  \
  
  A  natural   way  to resolve  this problem is to  assume 
  that the 10d action \rf{2}  is  missing some  ``boundary term''  that restores the 
 equivalence of its on-shell  value  with that of  the 5d   action   \rf{1}.  
 However,   such term  cannot be one of the   familiar  choices  like the Gibbons-Hawking-York (GHY) one 
 \ci{York:1972sj,Gibbons:1976ue}\foot{
 This term does not contribute in the case of AdS asymptotics.} 
   or 
 boundary terms that   may be  added to the 5d action \rf{1}  to make it  IR finite
  when  evaluated on a classical solution with \ads asymptotics  (see, e.g., \ci{Liu:1998bu,Henningson:1998gx,Arutyunov:1998ve,Emparan:1999pm,deHaro:2000vlm}).

An  important general point  is that   boundary or topological terms   may not be universal: they 
 may depend   on  a choice of vacuum (near which one expands in order to find an  effective action for fluctuations) 
or asymptotic boundary conditions. 
For example, in  the type IIB string   theory there  are  two maximally supersymmetric vacua   --  the  flat   space  R$^{1,9}$ 
 and \adss   \ci{Schwarz:1983qr}   -- that   have  different asymptotic symmetries.  
  The  corresponding effective actions     may, in principle,    contain   different   boundary  terms.
  
  In what follows   we will  be interested in the case   when  
  the topology of 10d space-time   is   that of  a   product $ M^5 \times \K^5$  where $M^5$ is   non-compact 
  and   $\K^5$   is  a compact space.  
We  will   suggest a  novel  5-form  dependent  ``topological''  term  that   should   be added to  the 10d action \rf{2} 
  to   restore  its   on-shell equivalence with the reduced  5d action \rf{0}.\foot{
    In the case of solution of 6d theory (obtained by compactification on $T^4$) 
     supported   by self-dual 3-form   flux one  will need  to add a topological term built out of $H_3$. 
    In the case of  the ${\rm AdS}_3 \times S^3\times T^4 $  solution  supported  by 5-form flux  discussed in  section 3 one will need  the same $F_5$-dependent  topological term.}

 \
 
Let us  stress again that   the  reason why   one 
 would like to understand the 10d origin of the on-shell value of the reduced action  like \rf{0}
is that it should  have a  string theory origin
(being related to  
  string  partition   function on  a 2-sphere). 
For example, the   tree-level   bosonic string  effective action
 may be   written as\foot{This   action   may be  reconstructed  also 
   from scattering  amplitudes near  asymptotically flat vacuum, with the  boundary term required for a  consistent definition of the 
    graviton/dilaton  S-matrix.}
\begin{align} 
&\la{7} S_D = S_{\rm bulk} + S_{\rm bndry}, \ \ \ \      \qquad  S_{\rm bulk}= \hat S_D  = \vk \int d^D x \sqrt{G}\ e^{-2 \p} \, \td \beta^\p \ , 
 \qquad \ \  \vk= {\te {2\ov \kappa^2_D \a'} } \ , 
 \\
&\la{8} \td \beta^\p = c_0 - \tfrac{1}{ 4 } \a' \Big( R + 4 \nabla^2 \p - 4 \del_\m \p \del^\m \p -  \tfrac{1}{2}{|H_3|}^2  \Big)  + \OO (\a'^2) \ , \ \ \ \ \ \
c_0 = \tfrac{1}{ 6 } (D-26) \, , \\
&\la{9}  S_{\rm bndry}= -\ha  \vk\, \a' \int d^{D-1} x \sqrt \gamma\,    e^{-2\p} ( K - 2 \del_n \p) = -\ha \vk\, \a' \int d^{D-1} x \sqrt \gamma\,  \nabla_a (e^{-2\p}  n^a) \ .
  \end{align}
Here the integrand   \rf{8} of the bulk part  is   proportional to the generalized   conformal anomaly coefficient $\td \beta^\p$
 and thus  must  vanish  on-shell\foot{Strictly speaking, this is true for   backgrounds 
   for which there is no source in the dilaton equation, cf. discussion of brane solutions in Appendix B.}
    not  only to first two  leading  orders \ci{Callan:1986jb,Tseytlin:1986ti}  but also to all orders in $\a'$ 
 \ci{Tseytlin:1986tt,Tseytlin:1987bz,Tseytlin:1988tv,Osborn:1989bu}.\foot{The same   conclusion was 
   reached 
   for the on-shell value of the closed bosonic string field theory action  \ci{Erler:2022agw}.}
 The boundary term \rf{9} 
 which is  a dilatonic   generalization \ci{Kazakov:2001pj} of the  standard GHY term 
    may,  in general,     produce   a non-zero  on-shell value  for     the total  action.
    
 Similar    remarks   apply   to the NS-NS part of the type IIB  superstring  effective action. 
   Note that the   boundary term  that should be added in general   to the bulk type IIB action \rf{2}
    (with the second-derivative dilaton term in  \rf{8}  integrated by parts
    and thus not automatically vanishing on  solutions with non-constant dilaton) 
  is given   by  \rf{9}   without  the $\del_n \phi$ term, i.e. 
 \be 
 \la{99}  S_{\rm bndry}= -  {1\ov \k_{10}^2} 
  \int d^{D-1} x \sqrt \gamma\,    e^{-2\p} \,  K \ . \ee
   As for 
 the R-R   terms in  the second line of \rf{2}, they   may    lead to additional non-trivial   boundary 
 contributions when evaluated on a  classical solution.\foot{For example,  $|F_3|^2$ term  
   reduces to a boundary  term upon use of the  field  equation  $\nabla_\m F^{\m\n\l} + ...=0$, cf. also \ci{Chen:2021dsw}. } 
   Given that  the   bulk  $|F_5|^2$  term vanishes identically upon use of the on-shell 
   self-duality condition,   an extra   $F_5$-dependent  contribution to 10d action   would be 
    required  to get a  non-zero contribution  for solutions  with   only    $F_5$-flux   being   non-zero. 
   This   new  term  should not change the  equations  of motion,  i.e.  it  should be a  ``topological''
     or ``boundary'' term.\foot{Note that the fact that particular topological  or  boundary terms  may   or may not be relevant 
    depending on boundary asymptotics of the fields  is not unfamiliar. 
   For example, the  GHY  boundary term  complementing the Einstein action  is relevant in   the 
   asymptotically flat space but   may not  be  contributing in  the AdS case  (e.g. it 
vanishes for the AdS Schwarzschild black hole   because the black hole correction to the AdS metric vanishes too rapidly at infinity
 \ci{Hawking:1982dh,Witten:1998zw}).
 } 
 
 \
 
 We shall  suggest such a topological  term in section 2. 
 In section 3 we shall compute the value of the full 10d action (containing the bulk  term \rf{2}, the boundary term \rf{99}   as well as the topological term) on the  extremal D3-brane solution and  its non-extremal   generalization. We shall  also   note the  non-zero value of  the topological term on   solutions describing  BPS intersections of two and four  D3-branes that  in the near-core limit reduce to   AdS$_3\times S^3 \times T^4$   and  AdS$_2\times S^2\times T^6$  backgrounds respectively. 
 Section 4 will contain some concluding remarks.  In Appendix A we shall argue that the presence of the same  topological term is suggested also  by gauge invariance requirement in the PST formulation
 \ci{Pasti:1996vs,DallAgata:1997gnw,DallAgata:1998ahf}  of type IIB  supergravity action. 
 In Appendix B we shall  discuss the  computation of the value  of the  10d action on  fundamental string, NS5-brane    and D5-brane solutions.

\def \pp {{\rm k}}

\section{Topological term} 

  While  the   obvious  guess for the 10d topological invariant  $\int F_5 \we F_5$  is identically zero, 
    a   non-trivial   candidate is possible   if  we assume that the 10d space has a particular topological  
  structure. 
  Namely,  let us specify to the 
    backgrounds   for which the 10d 
 space-time   is  a product $M^5 \times \K^5$  where    $M^5$ is  non-compact  (e.g.,  asymptotically AdS$_5$)     
  while $\K^5$ is compact  and a 
    similar factorization applies   to  the 5-form field strength  (for simplicity,  we shall     ignore  all other fields)
    \be \la{11} \rM^{10} = M^5 \times \K^5 \ , \qquad \qquad 
        F_5= F_{5M} \oplus F_{5\K}\ ,   \ee   
         and also its   potential $C_4= C_{4M} \oplus C_{4\K}$. 
 Then   consider the  following   ``topological'' term  
\be\la{10}   S_\top =   \g  \int F_{5M} \we F_{5\K}   \ .  \ee    
 As  $M^5$ is non-compact and $F_{5M}= d C_{4M}$ while $dF_{5\K}=0$  this term reduces to a boundary contribution
 and thus does not  affect  the  bulk equations of motion. 
 
Integrating    over the compact $\K^5$ then gives   
\be     S_\top= \g\,  q \int_{M}   F_{5M}\ , \ \   \qquad \qquad q =   \int_{\K}  F_{5\K}   \ .  \la{100} \ee
The  integral of a 5-form $F_{5M}$   is  effectively equivalent to an extra  $M^5$   volume   term.
Equivalently,  using  the on-shell condition of  selfduality of $F_5$ giving  $F_{5\K}= {}^*F_{5M}$   we conclude that 
$\rS_\top= \g\int F_{5M} \we {}^*F_{5M}  \sim  \vol(\K^5) \int_{M}  |F_{5M}|^2$, 
which again  produces, as is well known  \ci{Aurilia:1980xj,Duff:1980qv,Duncan:1989ug}, a  contribution to  
 5d   cosmological  term. 

 More  generally,  the  assumption of  simple ``5+5''  factorization of $F_5$   may be relaxed: 
 provided  $F_5$   can be split into an ``electric''   part (involving time differential)  and  its  dual magnetic part   the topological term 
   may be written as 
 \be\la{1000}   S_\top =   \g  \int F^{\rm (el)}_{5} \we F^{\rm (mag)}_{5}   \ , \qquad 
 \qquad    F^{\rm (mag)}_{5} = {}^* F^{\rm (el)}_{5} \ .   \ee  
 The 
  value of  the coefficient $\g$ in \rf{10},\rf{1000}
   required  to match  the  coefficient of the cosmological term in \rf{1} is\foot{Note that in our  notation
   (with Minkowski signature 10d metric)  for a general 5-form one has:
   
    \ \   
  $\int F_5 \wedge {}^*F_5=-  (5!)^2  \int d^{10} x \sqrt G\,   |F_5|^2  $.}
 \be  \g= -\frac{1}{4\, (5!)^2 \, \kappa_{10}^2 }   \ ,      \la{101}
\ee 
so that  the topological term in \rf{1000} takes the form 
\be 
S_\top =   -\frac{1}{4\, (5!)^2 \, \kappa_{10}^2 }   \int F^{\rm (el)}_{5} \we {}^* F^{\rm (el)}_{5}  
=   \frac{1}{4 \, \kappa_{10}^2 }   \int d^{10} x \sqrt G\,   |F^{\rm (el)}_{5} |^2\ . 
\la{000} \ee
The total 10d action is then  given by the sum of the bulk term \rf{2}, the new  topological term  \rf{10}  and   the 
 boundary term  \rf{99} 
 \be S_{10} = \hat S_{10}   + S_\top + S_{\rm bndry}   \ .\la{10001} \ee
 Note that  the $|F_5|^2$ term in  the bulk action \rf{2}     may be written (before imposing self-duality) 
as 
$\frac{1}{8 \, \kappa_{10}^2 }   \int d^{10} x \sqrt G\,   (|F^{\rm (el)}_{5} |^2  +  |F^{\rm (mag)}_{5} |^2)$.
Adding the topological term \rf{000}   corresponds effectively  to  reversing  the  sign of the magnetic part  in $|F_5|^2$, thus     doubling the  contribution of the electric part once going on-shell
 (the self-duality condition implies  $|F^{\rm (el)}_{5} |^2  = -  |F^{\rm (mag)}_{5} |^2$). 


Let us note that a   similar procedure  of inverting the sign of the  square of the ``electric components''
  of field strength in the action  was used also 
in  the discussion of  flux compactifications (cf. \ci{Polchinski:2010hw,Jockers:2004yj,Grimm:2008dq}). 
This may be interpreted  as implied by the ``democratic''  formulation of supergravity  \ci{Bergshoeff:2001pv}  with doubled number of RR   fields.\foot{For example, one may start with an action containing two unconstrained 5-form field strengths $F'_5$  and $F''_5$   
and consider      configurations in which $F'_5$
 has  electric    part  only, and  $F''_5$   the  magnetic part  only. 
 Dualizing  electric $F'_5$  will convert it to magnetic one and thus effectively double the total magnetic  contribution.}

As a result, the    value \rf{0}   of the tree-level  type IIB   action on the \adss 
  vacuum  solution 
  comes  entirely  from the topological term  \rf{10},\rf{101}:
  using    \rf{4} we get   
 \be \la{33} 
 S_{10}  \Big|_{{\rm AdS_5} \times S^5}= S_{\top} \Big|_{{\rm AdS_5} \times S^5}  =- \frac{1}{4  \, (5!)^2 \, \kappa_{10}^2} \int 
 F_{\rm AdS_5}\wedge F_{S^5}
 = - {4L^8\ov \kappa_{10}^2} \, { \rm vol (AdS_5) } \ , 
 \ee
 which is  the same  result  that    follows  from  the 5d action \rf{0}. 
 
 This   has straightforward generalization  to the case of ${\rm AdS_5} \times \K^5$   solutions   where $\K^5$ is an Einstein manifold 
 as in  \ci{Gubser:1998vd}:    instead   of \rf{0}  one gets $ S_5= \pp\,  N^2  \log ( \L \rr) , $  with    $\pp\equiv  {   {\rm vol}(S^5) \ov  {\rm vol}(\K^5) } $ and   $L^4 = 4 \pi \a'^2 g_s\,  \pp\, N$. 
  
 \
 
As we will    show  in Appendix A, the  same term \rf{10}   with  precisely the same coefficient \rf{101} 
is also required for  gauge invariance 
in the  PST formulation \ci{Pasti:1996vs,DallAgata:1997gnw,DallAgata:1998ahf}  of  the 10d supergravity   action   
where  the 5-form  self-duality  condition  follows from  the equations of motion. 

\ 

To provide further evidence  that adding the term \rf{10}  to the type IIB action \rf{2}  restores its on-shell equivalence 
with the  5d reduced action like \rf{1} 
let  us  consider the following $M^5 \times S^5$ 
ansatz   for the metric and  $F_5$ (with  its  self-duality condition relaxed
and all other fields   set to zero)  
\be\la{5}
ds^2_{10} = L^2 \big[ e^{-{10 \ov 3} \n(x)} g_{mn} (x) dx^m dx^n + 
e^{2 \n(x)} d \Omega_5^2\big] 
\ ,  \ \ \ \ \ \  \ \ \   F_5 =  4 L^{-1}  \big[\aa(x)\,  w_5 +  \bb\,  {\rw}_5\big]\ .   \ee
Here   $x=\{x^m\}$\ ($m=0, 1, ...4$), \  $w_5$   and $\rw_5$ are the  volume forms on  $M^5$ (with metric $g_{mn}$)  
 and $ S^5$  and we extracted the factors of the overall scale $L$. 
Following 
     \ci{Liu:1999kg} we  introduced the  warp factors   depending on a  ``fixed scalar'' $\n(x) $.\foot{The specific  dependence on $\nu$   in the metric
 is required to decouple $\n$ from the 5-d graviton;
this  generalizes  the graviton mode  decomposition  in  \ci{Gunaydin:1984fk,Kim:1985ez}  where 
$\n$ was  identified with  the zero mode  of the trace of the  perturbation of the metric of $S^5$.} 
The condition  $dF_5=0$ implies that $\aa=\aa(x)$ and $\bb=\const$. 
Then the $R-  \frac{1}{ 4}  |{{F}_5}|^2$    part of the 10d action \rf{2} 
compactified  on $S^5$     becomes 
 \begin{align}
 & \la{6}
\hat S_5 =  - {1 \ov 2 \ka_5^2} \int d^5 x \sqrt{g} \, 
\Big[ {R}_5 -  \tfrac{40 }{ 3} (\del_m \n)^2  - V(\nu) 
       + ... \Big]  \, , \\ 
& \la{66}       V(\nu) = L^{-2} \big(-  20 e^{-{16 \ov 3} \n}    -    4 \aa^2  e^{{40 \ov 3} \n}  +     4 \bb^2  e^{-{40 \ov 3} \n}  \big)
  \ . \end{align}
The 3  terms in the potential $V$ 
originate from
 the  scalar  curvature of $S^5$ 
and the $|F_5|^2$  term  in \rf{2} (cf.  \ci{Liu:1999kg,Maldacena:2003nj}).
Using  the on-shell self-duality  of $F_5$  that  gives  $\aa=  e^{-{40 \ov 3} \n} \bb$ 
we find that the last two terms   in the potential \rf{66} mutually cancel   and thus, as was already mentioned above, 
 we do not reproduce 
the  value of the cosmological constant in \rf{1}. 

If instead   one plugs the ansatz \rf{5} into the  10d  equations of motion for \rf{2}  (that imply  that 
$\bb^2=1, \  \aa=  e^{-{40 \ov 3} \n}$) and then reconstructs the corresponding
  effective action 
for the remaining 5d fields $g_{mn}(x)$ and $\nu(x) $  one  finds instead 
 the action \rf{6}   with the following potential \ci{Liu:1999kg}
\be \la{77}
  V(\n)= L^{-2} \big( -  20  e^{-{16 \ov 3} \n}     +     8   e^{-{40 \ov 3} \n}\big) \ .  \ee
  This potential has the  minimum  at $\nu=0$  and  where   it  reproduces 
   the cosmological   term  $12 L^{-2}$ in \rf{1}.
    Comparing to \rf{66},  the potential \rf{77}  has  the sign of the middle  $a^2$ term in \rf{66} effectively reversed 
     so that it doubles the coefficient of the  last $b^2$ term upon use of the on-shell condition  $\aa=  e^{-{40 \ov 3} \n}$.  
     
 This is     precisely what happens if we   add to \rf{6}  
 the contribution of the topological term \rf{10},\rf{101}  and then   use the   self-duality of $F_5$. 
 We conclude  that adding this term to the  type IIB action  ensures the   equivalence 
 between the 10d and 5d actions  not only for \adss    but also    for more general   solutions   of $M^5 \times S^5$ topology.

 \section{10d action   on  D3-brane solutions}

Let us  now generalize  the above discussion of the  on-shell value of the type IIB action \rf{2}  with   the topological term \rf{10}  added 
to the case of the extremal    and non-extremal  D3-brane   solutions  that also have  the product   topology  as in  \rf{11}.

  The extremal  D3-brane solution 
   is given by  \ci{Horowitz:1991cd,Duff:1991pea}
\begin{align}
&ds^2_{10} =h^{-{1/ 2}}(r)\, dy^{\m}dy_{\m} + h^{1/2}(r)(dr^2 +r^2d\Omega_{5}^2) \ ,  \qquad 
   h(r)=1+\frac{L^4}{r^4} \ ,  \ \ \ \  L^4 = 4 \pi \a'^2 g_s N , \la{34}\\
&\la{35} C^{(\rm el)}_{4}= \big[
 h^{-1}(r) -1\big]
 dt \wedge dy^1 \we dy^2 \wedge dy^3  \ , \quad \ 
F_{5} = F^{(\rm el)}_{5}+ F^{(\rm mag)}_{5}\ , \   \quad  F^{(\rm mag)}_{5} =  {}^* F^{\rm (el)}_5  \ ,  \ \\
& \la{36}
F^{(\rm el)}_{5}= \frac{4r^3L^4}{(r^4+L^4)^2}dt\wedge dy^1 \we dy^2  \wedge dy^3 \wedge dr \ , \qquad \ \ \ \ F^{(\rm mag)}_{5}={4}{L}^{-1}  {\rm w}_5 
\ .
\end{align}
Here $y^\m= (y^0\equiv t, y^1, y^2, y^3)$   are  coordinates  along the  D3-brane     and ${\rm w}_5 = \sqrt{g_{_{S5}}}dz^{5}\wedge ... \wedge dz^{9}$ is the volume  form of $S^5$.  The near-core  limit
 $h\to \frac{L^4}{r^4}$ corresponds to 
 the  \adss  case. 


As discussed in the Introduction, the bulk part of the  type IIB action \rf{2}   has zero on-shell value  
(once again, the  self-duality of $F_5$  implies $|F_5|^2=0$ and thus also $R=0$). 
A 
    non-trivial  contribution   may come from the topological term \rf{10} 
     and also from  the GHY   boundary term \rf{99} that  may  be non-vanishing  in this   asymptotically flat   case. 
From \rf{10},\rf{100}   we find    (cf. \rf{33})\foot{If  
 we focus on  the near-core limit  ($r \ll L$) of  \rf{34},\rf{35}
we  get the same expression 
as in \rf{33}   with the volume of AdS$_5$ written in Poincare coordinates.} 
\begin{align}
 S_{\top} \Big|_{\rm D3 } &=- \frac{1}{4  \, (5!)^2 \, \kappa_{10}^2} \int F_{5M}\wedge F_{5\K}
 = - \frac{1}{4  \, (5!)^2 \, \kappa_{10}^2} \int  F^{(\rm el)}_{5} \wedge  F^{(\rm mag)}_{5}\no \\
 &= - \frac{{\rm vol(S^5)}}{2\kappa^2_{10}} \int_{0}^{\infty} \frac{ dr\   8 L^8 r^3}{(r^4+L^4)^2}\ \int d^4 y
  = - \frac{{\rm vol(S^5)}}{\kappa^2_{10}}   L^4   \int d^4 y = -\tfrac{1}{2} N \mu_{3} \int d^4 y \la{37}
  \ .  \end{align} 
  Here  
  \be\la{73}  
    \mu_{3} = { 2{\rm vol(S^5)}   L^4 \ov N \kappa^2_{10}} =  { 1\ov (2 \pi)^3 g_s \a'^2}
  \ee  is  tension of a  unit-charge 
    D3-brane (cf. footnote \ref{f1}) and $\int d^4 y$ is the  integral over the D3  world volume directions. 
  Compactifying $(y^1,y^2,y^3)$  on a torus  with volume  $\V_3$ 
    we  get 
  \be \la{74}
    S_{\top} \big|_{\rm D3 } =  - \ha N  M_3  \int dt \ , \qquad     M_3 = \mu_{3} \V_3\ , \qquad   \V_3= \int d^3 y   \ ,  \ee
where $M_3$  is    the  mass  of a single D3-brane.

  The GHY  boundary term  \rf{99}  (that did  not contribute in  the \adss case) 
    happens  to  give the same result  as in \rf{37}  
 (here  the  asymptotic  boundary is  at $r=\infty$)\foot{Here and below  when 
  evaluating the boundary term \rf{99} 
  we neglect  contributions  that are independent of  the parameters of  the solution.
  }  
\begin{equation}\la{38}
S_{\rm bndry}\Big|_{\rm D3 } =-\frac{{\rm vol(S^5)}}{\kappa_{10}^2}   \frac{L^4}{1+{L^4\ov r^4}}\Big|_{ r \rightarrow \infty }  \int d^{4}y  =-\tfrac{1}{2} N \mu_{3} \int d^{4} y \ .
\end{equation}
Then  the on-shell  value of the   10d   action \rf{10001}
 on  the D3-brane solution  is given by 
\begin{equation}\la{39}
S_{10} \Big|_{\rm D3 } =(S_{\top} +S_{\rm bndry} ) \Big|_{\rm D3 }
=- N \mu_{3}\int d^4 y \ . 
\end{equation} 
 In addition,  one  may  consider  the value of the 
  D3-brane  source  action    that  provides  the delta-function   
 in the   equation for the  harmonic  function 
  $h(r)$
\be \la{377}
S_{\rm source}
=-N\mu_{3} \int d^{4} y \sqrt{G_{4}} +  
 N\mu_{3} \int 
  C_{4} \ . 
\ee
 More generally, considering this  as an action  of a static probe D3-branes placed at distance $r$ parallel to the source branes at $r=0$
one  finds  from \rf{34},\rf{35}  that 
the $h^{-1}$  factors from the two terms in \rf{377}   cancel each other\foot{This is,  of course,   
 a manifestation of  the BPS condition of the   vanishing   force, see, e.g.,  \ci{Tseytlin:1996hi}.}
leaving simply \be\la{773}  S_{\rm source} \Big|_{\rm D3 } = - N \mu_{3}\int d^4 y \ee 
coming 
from the $-1$   in $C_4$ in \rf{35}.
 This is equal to  the free  brane  action at $r=\infty$   and the same  expression is  thus also  at $r\to 0$.

 As a result, the total action  on D3-brane solution  is given by 
 \be  \la{001}    
 S_{\rm tot} \equiv  S_{\rm bulk}  + S_{\rm top} +  S_{\rm bndry} + S_{\rm source}  \ , \qquad \ \ \ \ 
 S_{\rm tot} \Big|_{\rm D3 } =  - 2 N \mu_{3}\int d^4 y\ . 
 \ee
Similar  computations  of  the value of 10d  action on some  other p-brane  solutions  are presented 
in Appendix B. 


Next, 
let us consider the non-extremal (black)   D3-brane solution \ci{Horowitz:1991cd}  generalizing \rf{34}--\rf{36}\foot{We use the same parametrization as in \ci{Tseytlin:1998cq}.}
\begin{align}
&ds^2_{10} =h^{-{1/ 2} }(r)\big[-f(r)dt^2+ dy^{i}dy^{i}\big]  + h^{1/ 2}(r)\big[f^{-1}(r)dr^2 +r^2d\Omega_{5}^2\big] \ , \la{340} \\
&\la{341} h(r)=1+\frac{\M^4}{r^4} \ , \qquad f(r)=1-\frac{r_{0}^4}{r^4} \ , \qquad \M^4=\sqrt{L^8+\tfrac{1}{4}r_{0}^8}-\tfrac{1}{2}r_{0}^4 \ , \\
&\la{350} C^{(\rm el)}_{4}= \sigma  \, [h^{-1}(r)-1]  dy^{0}\wedge ... \wedge dy^{3}  \ , \qquad 
 \sigma\equiv {L^4\ov \M^4}=  \sqrt{1+\tfrac{r_{0}^4}{\M^4}} \ ,  \\
& \no
F^{(\rm el)}_{5}= \frac{4\sigma \M^4r^3}{(r^4+\M^4)^2}dy^{0}\wedge ... \wedge dy^{3} \wedge dr \ , \qquad F^{(\rm mag)}_{5}={4\sigma}{\M}^{-1} {\rm w_5} \ ,     \qquad F_{5} = F^{(\rm el)}_{5}+    F^{\rm (mag)}_5 \ ,   
\end{align}
where $L$ is the same  as in \rf{34}. 
We shall consider   this solution  for $r_0 \leq r < \infty$   and     should  not  introduce  an explicit  brane source. 

 The value of the  topological  term \rf{100}
is found   as in \rf{37}
\begin{align}
 S_{\top} \Big|_{\rm  black \, D3}  
& = - \frac{1}{4  \, (5!)^2 \, \kappa_{10}^2} \int  F^{(\rm el)}_{5} \wedge  F^{(\rm mag)}_{5}
=- \frac{{\rm vol(S^5)}}{2\kappa^2_{10}}\sigma^2 \int_{r_0}^{\infty} \frac{ dr\   8 \M^8 r^3 }{(r^4+\M^4)^2} \int d^4 y\no \\
& =- \frac{{\rm vol}(\rm S^{5})}{\kappa^2_{10}} {  \M^4} \int d^4 y 
  \la{370}
  \ .  \end{align} 
  Once again we see that the topological term gives a non-trivial contribution to the action. 
  
  The expression \rf{370}  may be written  also as 
   \begin{align}
   S_{\top} \Big|_{\rm  black \, D3} &=\tfrac{1}{2}N\mu_{3}\, C^{(el)}_{4}(r_{0})\int d^4 y \ , \la{613} \\
      C^{(\rm el)}_{4}(r_{0})&=-\frac{\sigma \tilde{L}^4}{r_{0}^4+\tilde{L}^4} \ , \qquad \ \ \  \ \ N\mu_{3} = { 2{\rm vol(S^5)}   L^4 \ov  \kappa^2_{10}}={ 2{\rm vol(S^5)} \sigma  \tilde{L}^4 \ov \kappa^2_{10}}\ , 
  \end{align}
  i.e. is proportional  to a product of  the electric potential $C^{(\rm el)}_{4}$
    at the horizon and the black D3-brane charge.  
   This is  analogous  to what one finds in the case of the Reissner--Nordstrom black hole  \ci{Gibbons:1976ue}.\foot{In the context of black brane thermodynamics 
  the topological term  will  thus  contribute to the part of the ``thermodynamic potential''
    related to the product of the  chemical potential and the corresponding conserved charge.}
   
  
The calculation of the   asymptotic  $r\to \infty$ boundary  GHY term \rf{99}    here gives   (cf. \rf{38}) 
\begin{align}
S_{\rm bndry}\Big|_{\rm   black \, D3 }=&-\frac{ {\rm vol(S^5)} }{\kappa_{10}^2} \Big[\M^4\frac{1-{r^4_{0}\ov r^4} }{1+{\M^4\ov r^4}}+3r_{0}^4\Big]_{r\rightarrow  \infty} \int d^{4}y  
=-\frac{   {\rm vol(S^5)} }{\kappa_{10}^2}   \, (\M^4+3r_{0}^4)  \int d^{4}y \ .     \la{333}
\end{align}
As the bulk 10d action \rf{2} is  again   vanishing, 
   the total action  \rf{10001}   computed on the  non-extremal  D3-brane solution then follows by combining \rf{370}  and \rf{333} 
\begin{align}\la{390}
S_{10} \Big|_{\rm black \, D3} =&(S_{\top} +S_{\rm bndry} ) \Big|_{\rm  black \, D3 }
=-\frac{2\, {\rm vol(S^5)} }{\kappa_{10}^2}  \, (\M^4+\tfrac{3}{2} r_{0}^4)  \int d^{4}y \no  \\
=&-\frac{2\, {\rm vol(S^5)} }{\kappa_{10}^2}   \big( \sqrt{L^8+\tfrac{1}{4}r_{0}^8} +  r^4_0 \big)
 \int d^{4}y  
     \ . 
\end{align}
The same   result   should  be  found   by first compactifying on $S^5$, finding the reduced   5d action  generalizing  \rf{1} and then evaluating it on the  corresponding 5d black  brane  solution.\foot{For comparison
 with the  extremal  case  \rf{74}  let us note that  the ADM  mass of  black D3-brane is given by 

$\widetilde M_3= \mu_3 {\rm V_3}  {\M^4+{5\ov 4} r_{0}^4 \ov L^4} 
= M_3 \big[ \sqrt{1+\tfrac{r_{0}^8}{4L^8}} + \tfrac{3r^4_0}{4L^4} \big].$ }



\iffa
The ADM mass of black D3-brane is proportional to ${\sigma}{\M}^{-1}= {\M}^{-1}  \sqrt{1+\tfrac{r_{0}^4}{\M^4}}$
so why it is  not  appearing in \rf{390} ???
{\bf express this  in terms of  mass of   D3 defined as mass  of black hole; cf.  book by Ortin for example   
or \ci{Klebanov:1996un} }
\fi

\

Similar   discussion  can be repeated   for the  type IIB  solutions 
describing   BPS intersections  of D3-branes  -- 
D3$\perp$D3 \ci{Tseytlin:1996bh}  and D3$\perp$D3$\perp$D3$\perp$D3 \
\ci{Klebanov:1996mh}. In the near  core limit  they  reduce (in the extremal case)
 to 
AdS$_3\times S^3\times  T^4$   and AdS$_2\times S^2\times  T^6$    backgrounds  respectively.
Here the bulk part of type IIB  action  is  again  vanishing,   with  possible 
non-zero contribution  coming  from  the 
topological  term  \rf{1000} defined in terms of 
\be \la{5335}   F^{(\rm el)}_{5}= dC^{(\rm el)}_{4} \ , \qquad F^{(\rm mag)}_{5} =  {}^* F^{\rm (el)}_5 \ , \qquad 
 F_{5} = F^{(\rm el)}_{5}+ F^{(\rm mag)}_{5}\ , \ee
 and  also  the GHY  term (in the case of the full asymptotically flat solution). 

The D3$\perp$D3    solution is the following generalization of the D3   background \rf{34}--\rf{36}:
\begin{align}
ds^2_{10} = &(h_1 h_2)^{1/2} \Big[   (h_1 h_2)^{-1}  \, (-dt^2 + dy_1^2) 
+ h^{-1}_1 ( dy_2^2 + dy_3^2)  + h^{-1}_2 ( dy_4^2 + dy_5^2) \no 
\\ & \qquad \qquad \qquad \qquad \ +    dr^2 +r^2d\Omega_{3}^2\Big] \ ,\qquad \qquad 
   h_i=1+\frac{L_i^2}{r^2} \ , \no \\
\la{335} C^{(\rm el)}_{4}= &\big[ h^{-1}_1 -1\big]
 dt \wedge dy^1 \we dy^2  \wedge dy^{3}  +   \big[ h^{-1}_2 -1\big]
 dt \wedge dy^1 \we dy^4  \wedge dy^{5}    \ . 
\end{align}
Here $( y^1, y^2, y^3)$     and $(y^1, y^3, y^4)$ are   spatial coordinates  along the two 
 D3-branes intersecting over $y^1$ direction.  In the   near-core limit $h_i\to {L_i^2\ov r^2}$ this  background reduces 
 to AdS$_3\times S^3\times  T^4$    with  $ds^2_{\rm AdS_3} = {r^2 \ov L^2} ( - dt^2 + dy_1^2)   + {L^2 \ov r^2} dr^2$, \ $ds^2_{S^3} = L^2 d\Omega_{3}^2$ (where $L^2= L_1 L_2$)  
 and $ds^2_{T^4}=  {L_2\ov L_1} ( dy_2^2 + dy_3^2) +  {L_1\ov L_2} ( dy_4^2 + dy_5^2) $.
 
Note that   here  $F_5$  does not have a simple  5+5  decomposition 
so     the topological term term  is   defined  by \rf{1000} or, equivalently, \rf{000}.
Computing it   gives 
       a non-zero  value consistent with  the one of the  dimensionally reduced  3d  analog of the action \rf{1}  that admits AdS$_3$ as  its solution.  Explicitly,  in the AdS$_3\times S^3\times  T^4$   limit we find that 
   $S_{\rm top} = -{2\ov \kappa_{10}^2} \vol ({\rm AdS}_3) \, \vol(S^3) \, \vol(T^4)$
   (where   we did not extract the dependence on the scale $L=\sqrt{L_1 L_2}$).

 Similarly, the  four D3-brane solution  is given by 
 \begin{align}
ds^2_{10} = &(h_1 h_2   h_3 h_4)^{1/2}  \Big[ -   (h_1 h_2   h_3 h_4)^{-1}
\, dt^2 
+ (h_1 h_2 )^{-1}   dy_1^2 +  (h_1 h_3 )^{-1}     dy_2^2   + (h_1 h_4 )^{-1}    dy_3^2 
 \no 
\\ & +    (h_2 h_3 )^{-1}    dy_4 ^2  +   (h_2 h_4 )^{-1}    dy_5 ^2  +   (h_3 h_4 )^{-1}    dy_6 ^2   +    dr^2 +r^2d\Omega_{2}^2\Big] \ ,\qquad
   h_i=1+\frac{L_i}{r} \ ,\no \\
\la{3355} C^{(\rm el)}_{4}= &\big[ h^{-1}_1 -1\big]
 dt \wedge dy^1 \we dy^2  \wedge dy^{3}  +   \big[ h^{-1}_2 -1\big]
 dt \wedge dy^1 \we dy^4  \wedge dy^{5}  \no\\
 &+ \big[ h^{-1}_3 -1\big]
 dt \wedge dy^2 \we dy^4  \wedge dy^6  +   \big[ h^{-1}_4 -1\big]
 dt \wedge dy^3 \we dy^5  \wedge dy^6\ . 
\end{align}
This  reduces to AdS$_2\times S^2\times  T^6$    with the 6-torus   formed by $(y_1, ..., y_6)$. 
Here  the topological term \rf{1000},\rf{000}  produces again  a non-zero contribution to 10d action.

\section{Concluding remarks}
Depending on   topology  of  space-time or asymptotic boundary  conditions,  
the 10d  supergravity action  (or, more  generally, string effective action)
   may need to be supplemented  by particular  boundary or ``topological'' terms 
 specific to a   type of backgrounds  considered. 

Here we considered the case of ${\rm M}^{10} = M^5 \times X^5$   with  5-form   flux  and   showed  that 
adding the ``topological'' term  
\rf{10} or \rf{1000}
 to the bulk  type IIB  action \rf{2} restores its  equivalence with the 5d reduced  action 
(obtained  via equations   of motion  by compactifying on $X^5$). This  
 leads to  consistent  on-shell  values   of  the full 
 10d  action  (e.g.,   for  ${\rm AdS}_5 \times  X^5$ or D3-brane solution). 
Similar  terms  
 are to be added in   cases of  other topologies, e.g.,    $\int_6  F_{3M} \we F_{3X} $  for 
   ${\rm M}^{10} = M^3 \times X^3 \times  T^4$.\foot{Note  also  that  an  analogous example 
    is  found in the case of the extremal  dyonic  black hole    in 4d Einstein-Maxwell  theory. Here  the  AdS$_2 \times S^2$ vacuum  is supported by $F_2 = F_2^{\rm (el)} + F_2^{\rm (mag)}$
   and   the on-shell value  of the action is  zero ($R_4=0, \ |F_2|^2=0$).   Adding  the standard  topological term $\int F_2 \wedge F_2$ 
   then produces  a cosmological term in the  effective 2d action.
   This example is  related to the near-core limit of the  four D3-brane background discussed in the previous section. } 

String theory origin of  the term \rf{10}  and whether  it  may receive $\a'$ corrections  remains to be  understood. 
One  particular   case when the contribution of this term  may be important
 is the computation of $\a'$ corrections  to  near-extremal  D3-brane  entropy as in \ci{Gubser:1998nz,Pawelczyk:1998pb}.


\section*{Acknowledgments}
We are   grateful   to  M. Beccaria, E. Buchbinder, S. Giombi, T. Grimm,  K. Mkrtchyan,  J. Russo, K. Stelle    and T. Wiseman 
for  useful comments.
SAK  acknowledges  support  of  BASIS Foundation.  
AAT was supported by the STFC grant ST/T000791/1. 

\bigskip
\newpage

\appendix

\section{Topological    term in   $F_5$  action  on  $M^5 \times \K^5$   
from  PST formulation } \label{A}
\def\theequation{A.\arabic{equation}}
\setcounter{equation}{0}

In the  PST  formulation  \ci{Pasti:1996vs,DallAgata:1997gnw,DallAgata:1998ahf}  of the 5-form action
 the condition of self-duality 
is derived from an action. This is achieved by introducing 
  an extra  scalar field $a(x)$  along with extra gauge invariance 
   so that the number of dynamical  degrees of freedom is unchanged. 
    For  a  closed  5-form  $F_5$ 
     let   us 
 consider  the following action:\foot{As usual, $\mathfrak{i}_{v}$ 
 denotes the contraction of a differential form with a vector field, obtained from the coefficient  of 
  1-form $v$ by raising the index with the help of the metric. In \rf{21}   we ignore an overall normalization factor.}
  \be
\la{21} S_{\PST}= \int( F_{5}\wedge {}^*F_{5}+\mathfrak{i}_{v}\mathcal{F}\wedge^{*}\mathfrak{i}_{v}\mathcal{F}) =-\int 2v\wedge F_5\wedge\mathfrak{i}_{v}(F_5-{}^*F_5) \ , \qquad \mathcal{F}\equiv  F_{5}-{}^{*}F_{5}\ . 
\ee
We assume that 
 $F_{5}$ can be expressed locally as $F_{5}=dC_{4}$ (we ignore  all the other fields that may 
 contribute to $F_{5}$ in \rf{3}). \  $v= v_\m dx^\m$ is defined in  terms of 
   a scalar $a(x)$ as 
\be
v_\m =\tfrac{1}{\sqrt{-|\partial a|^2}} \del_\m a\ , \qquad \qquad  v^\m v_\m =-1 \ . \la{220}
\ee
The variation of \rf{21}  over $C_4$ and $a$ then  leads  to equations   that imply the self-duality 
condition $F_5={}^*F_5$. The dependence on the scalar $a$   drops out of the equations of motion.

The reason for this  is that 
apart from the standard  gauge symmetry of a 4-form potential $C_4\to C_4 + d \varepsilon_3$, 
the action \rf{21} is invariant (up to boundary terms, see below)  under the following  gauge  transformations 
\begin{align}
&\la{24} \delta_{\eta} a  = \eta\, , \qquad\ \ \ \  \delta_{\eta} C_{4}=-\tfrac{1}{\sqrt{-|\partial a|^2}}{\frak{i}}_{v}(F_5-{}^*F_5)\, \eta \, , \\
   &\la{23}  \delta_{\xi} a = 0\,  ,\ \ \ \  \qquad \delta_{\xi} C_{4} = \xi_{3} \wedge da \, , \ \ \ \qquad \delta_{\xi} F_{5} = 
   d\xi_{3} \wedge da \ .    
\end{align}
Here  the  scalar $\eta(x) $   and the 3-form $\xi_{3}(x)$    are  the 
gauge parameters.  The first   symmetry \rf{24} implies that $a$ is a pure gauge field.
The second      is effectively reflecting the fact that the number of degrees of freedom of $C_4$ 
is halved  on-shell (where  $F_5$  becomes self-dual).

Let us consider the  variation of the action under arbitrary  $\delta C_{4}$ and $\delta a$: 
 \begin{align}
    \la{28} \delta S_{\PST} &=  -\int \tfrac{ 2v}{\sqrt{-|\partial a|^2}}\wedge d \delta a \wedge\frak{i}_{v}\F 
    \wedge\frak{i}_{v}\F -\int 4v\wedge \delta F_{5} \wedge \frak{i}_{v}\F -\int 2F_{5}\wedge \delta F_{5} \no \\
    &=  -\int 2\delta a \ d\big[\tfrac{v}{\sqrt{-|\partial a|^2}}\wedge\frak{i}_{v}\F \wedge\frak{i}_{v}\F\big] -\int 4 \delta C_{4}\wedge d\big[v\wedge \frak{i}_{v}\F\big] \no \\
    &\ \ \ +\int_{\partial}\big[ \tfrac{2  v}{\sqrt{-|\partial a|^2}}\delta a \wedge\frak{i}_{v}\F\wedge\frak{i}_{v}\F + 4\delta C_{4}\wedge v \wedge \frak{i}_{v}\F\big]-\int 2F_{5}\wedge \delta F_{5} \ . 
 \end{align}
 Assuming that $\delta C_{4} = 0$ at the boundary, 
the resulting equations of motion may be written as:
\begin{align}
   \la{29} &\delta a : \ \ \ \ d\Big[\tfrac{v}{\sqrt{-|\partial a|^2}}\wedge \frak{i}_{v}(F_{5}-{}^{*}F_{5}) \wedge \frak{i}_{v}(F_{5}-{}^{*}F_{5})\Big]=0 \ , \\
    \la{290} &\delta C_{4} : \ \ \ \ d\Big[v\wedge \frak{i}_{v}(F_{5}-{}^{*}F_{5})   \Big]=0\ . 
\end{align}
Under the  transformation  \rf{23} the expression in brackets   in \rf{290} changes as:
\be
\delta_{\xi}\Big[v\wedge \frak{i}_{v}(F_{5}-{}^{*}F_{5})\Big]=-\delta_{\xi}F_{5} = -d\xi_{3}\wedge da \ , 
\ee
so that \rf{23}  is a symmetry of \rf{290}.  Furthermore, using \rf{220}  we may  choose 
such $\xi_{3}$ that $\frak{i}_{v}(F_{5}-{}^{*}F_{5})=0$.  Then  
\be
F_{5}-{}^{*}F_{5} = - v\wedge \frak{i}_{v}(F_{5}-{}^{*}F_{5}) + {}^{*}\big(v\wedge \frak{i}_{v}(F_{5}-{}^{*}F_{5})\big)=0 \ . 
\ee
Therefore, the symmetry \rf{23}   makes all solutions of \rf{290} 
 equivalent to the self-dual solution $F_{5}={}^{*}F_{5}$  (and all of them  lead to the vanishing on-shell value of $S_{\PST}$).  
 
Under \rf{23} 
 the integrand of \rf{21}  changes as:\foot{Here $ v\wedge\mathfrak{i}_{v}\delta_{\xi} F_{5}\wedge F_{5}=-\delta_{\xi} F_{5} \wedge F_{5} + v\wedge \delta_{\xi} F_{5} \wedge \mathfrak{i}_{v} F_{5}$  and 
 $ v\wedge \mathfrak{i}_{v}{}^{*}\delta_{\xi} F_{5}\wedge F_{5} = v \wedge {}^{\ast}(\delta_{\xi} F_{5}\wedge v)\wedge F_{5} =-\delta_{\xi} F_{5} \wedge v \wedge  \mathfrak{i}_{v}{}^{\ast}F_{5}$.}
\begin{align}\la{266}
\delta_{\xi}\mathcal{L}=&-2\Big[v\wedge \mathfrak{i}_v(\delta_{\xi} F_{5} - {}^{\ast}\delta_{\xi} F_{5})\wedge F_{5}+v\wedge\mathfrak{i}_{v}(F_{5}-{}^{\ast}F_{5})\wedge \delta_{\xi} F_{5}+v\wedge \mathfrak{i}_v(\delta_{\xi} F_{5} - {}^{\ast}\delta_{\xi} F_{5})\wedge \delta_{\xi} F_{5}\Big] \ .
 \end{align}
Using that 
$\delta_{\xi}F_{5} \wedge da =0$, the   variation of the action \rf{21}   may be written as  
\be 
\la{27}\delta_{\xi} S_{\PST} = -2\int F_{5} \wedge \delta_{\xi} F_{5} \ .
\ee
This vanishes if 10d space has no  boundary (as $dF_5=0$  we have  $F_{5} \wedge d\xi_{3}\wedge da=
 - d(F_{5} \wedge\xi_{3}\wedge da)$) but  otherwise  produces a boundary term.

Let us now assume as in \rf{11}   that  the 10d space  has a product structure, i.e.   $\rM^{10}=M^5 \times \K^5$ where $\K^5$ is a compact Euclidean space with no boundary 
 while $M^5$ (with Minkowski signature metric)  may be non-compact,  and also that a 
  similar factorization applies to the 4-form potential 
and the parameters of the  transformations in  \rf{23}, i.e. 
  \be \la{22} C_4= C_{4M} \oplus C_{4\K}\ \ , \qquad 
        F_5= F_{5M} \oplus F_{5\K} \ , \qquad  \delta_{\xi} C_4=\delta_{\xi} C_{4M} \oplus \delta_{\xi}C_{4\K} . \ee
In this case,    \rf{27}   takes the form 
\be\la{230}
\delta_{\xi} S_{\PST} = -2\int F_{5\K} \wedge  \delta_{\xi} F_{5M} =  2\int   \delta_{\xi} F_{M} \wedge F_{5\K}  =
   2\int_{\K} F_{5\K} \int_{M} \delta_{\xi} F_{5M} \ ,
\ee
where we used that $ \delta_{\xi} F_{5\K}$ is exact  so its integral over $\K^5$  vanishes. 
The  integral 
$\int _{M} \delta_{\xi} F_{5M}= \int_{\del M} \xi_3 \wedge d a $ 
depends on   the boundary  values 
of the gauge parameter $\xi_{3}$ and the scalar field $a$. If these are non-trivial     and if  $F_5$ 
has a  non-trivial  value of the ``magnetic''  charge  $\int_{\K} F_{5\K}\neq 0$, 
then the variation \rf{230} may be non-zero.

A  way to  maintain   the  invariance  of the action    \rf{21} under \rf{23}  is to add  to \rf{21}  the   topological   term    defined in  \rf{10} 
\be
\la{25} S_\top= -  2\int_{M} F_{5M}  \wedge F_{5\K}     = - 2 \int_{\K}  F_{5\K}\,  \int_{M} F_{5M}  \ .
\ee
The variation of this term under 
 the gauge transformation \rf{23}  will then   cancel  the   change \rf{230} of the PST action. 
 Assuming $F_{5M} = d C_{5M}$  is valid  globally on $M^5$,  the term  \rf{25} 
  may be expressed as an integral over the boundary  $\del M^5 \times \K^5$ 
   and thus  does not affect  the equations of motion for $F_{5}$.
   Let us note  that a similar argument   suggesting to add  the term \rf{25} to maintain gauge invariance can be given  \ci{Mkrtchyan:2022xrm} also in the formulation of self-dual $F_5$  field  suggested in \ci{Mkrtchyan:2019opf}.

     Using that  the equations of motion for \rf{21}  imply  the  self-duality  of  $F_5$, i.e. 
      $F_{5M}={}^{*}F_{5\K}$,  the  on-shell   value of \rf{21} plus  \rf{25}  may be written also as 
\be
\la{26}   (S_{\PST} +    S_{\top})\Big|_{F_5={}^{*}F_5}= S_{\top}\Big|_{F_5={}^{*}F_5}
 =2\int F_{5\K} \wedge {}^{*}F_{5\K} = - 2\int F_{5M} \wedge {}^{*} F_{5M} \ .
\ee
Replacing the $|F_5|^2$ term in the 10d  action  \rf{2}   by \rf{21}   one gets the corresponding  PST analog of the 
type IIB  action  to which now  we should add also \rf{25}   with the corresponding coefficient    being as in  \rf{101}.
It is interesting to  note   that the condition of the symmetry under \rf{23}  fixes also the  relative coefficient   between the kinetic 5-form term \rf{21} and the Chern-Simons type   term ($\int B_2 \we F_3 \we F_5$ in \rf{2})  in the resulting    version of   type IIB action  \ci{DallAgata:1997gnw,DallAgata:1998ahf}.


\section{10d  action  on F1, NS5 and D5   brane solutions } \label{B}
\def\theequation{B.\arabic{equation}}
\setcounter{equation}{0}

For comparison   with  the  case of the D3-brane   solution discussed in section 3    here we will discuss 
the values  of the 10d  action    \rf{001} on the   fundamental string, NS5-brane  and D5-brane extremal 
solutions. 
In these   cases $F_5=0$ so the topological term  \rf{10} will not play a role. 
These  $p$-brane  solutions  are  supported by sources  given  by  the corresponding brane   actions 
 that have the structure (see, e.g.,   \ci{Simon:2011rw,Ortin:2015hya}) 
\be \la{B1}
S_{\rm source}=- NT_{p} \int d^{p+1} y \, e^{-q \phi}  \sqrt{G_{p+1}} + N T_{p}\int 
A_{p+1} \ ,
\ee
where $T_p$ is a  tension of a single brane   and $A_{p+1}$  is the corresponding  NS-NS or R-R   potential.
The dilaton coupling  constant  is $q=0, 2,$ and 1  for  F1, NS5 and D5  cases respectively. 
\iffa 
where $y^u$ ($u=0,..., p$)  are  coordinates of world volume of p-brane, $g_{p}$ is induced metric on p-brane. The first term is just a Nambu-Goto part of the action, where $f(\phi)$ corresponds to the interaction with the dilaton field. The second term describes interaction of a p-brane with p+1-form potential that could be either electric (with the fields from  R-R sector or with the KR field) or magnetic (with the dual fields). Note that in order to possess the gauge invariance $A_{p+1}\rightarrow A_{p+1} +d\alpha$ of the action, one may consider the case of compactified spacial coordinates of a p-brane and require the gauge parameter to be zero at the time infinities $\alpha(+\infty)=\alpha(-\infty)=0$. The total action for a p-brane consists of the following parts
\fi 
The total action  will   be 
\be\la{b2} 
    S_{\rm tot}=S_{\rm bulk}+S_{\rm bndry}+S_{\rm source} \ , \ \ \ \ \  \ \ \ \ S_{\rm bulk} = \hat S_{10} \ , 
\ee
where   the bulk   part  is given by   \rf{2}  and the boundary one by \rf{99}. 

The F1 solution \ci{Dabholkar:1990yf}    is  electrically charged with the respect to the  $B_{2}$  field  ($T_1 = {1\ov 2 \pi \a'}$) 
\begin{align}
    &ds^2=H^{-1}(r)(-dy^2_{0}+dy^2_1)+dx^{a}dx^{a} \ , \qquad H(r)=1+\frac{Q}{r^6} \ ,  
     \qquad   Q=  { N  T_1 \kappa_{10}^2 \ov 3 \vol(S^{7})}= 32 N \pi^2 \a'^3  g^2_s  \ ,      \no\\
    & B_{2}=\big[ H^{-1}(r)-1 \big] \, dy^{0}\wedge dy^{1}  \ , \qquad e^{2\phi}=H^{-1}(r)    \ .    
    \la{bb4} 
\end{align}
Substituting this  solution into \rf{b2}    we find
\begin{align}
    &S_{\rm bulk}\Big|_{\rm F1 }=S_{\rm bndry}\Big|_{\rm F1 }=0 \ , \qquad S_{\rm source} \Big|_{\rm F1 }=
    -  N T_1 \int d^2y \ , \qquad S_{\rm tot}\Big|_{\rm F1 }=   -  N T_1 \int d^2y  \ . \la{bb5}
\end{align}
The magnetic dual  of F1-brane  --  the NS5-brane  solution \ci{Callan:1991dj} -- may be considered as electrically charged 
  with respect to  the dual field $\widetilde{B}_{6} $: \  $  d\widetilde{B}_{6}=e^{-2\phi}\, {} ^{\ast}H_{3}$,  i.e. 
  \be
S_{\rm source} =-NT_{5} \int d^{6}y \, e^{-2\phi}\sqrt{G_{6}}+NT_{5}\int
\widetilde{B}_{6} \ , \qquad \ \  T_5 = { 1\ov ( 2 \pi)^5 \a'^3 g^2_s} \ . \la{bb6}
\ee
The  corresponding  background is ($\m=0, ..., 5; \ a=6,7,8,9$; $r^2 = x^a x^a $) 
\begin{align}
&d{s}^2=\eta_{\m\n} dy^{\m}dy^{\n}+H(r)\, dx^{a}dx^{a}  \ , \qquad H(r)=1+\frac{Q}{r^2} \ , \qquad Q=   { N T_5\k_{10}^2\ov \vol(S^3)} =\a' N \ , \no \\
&\widetilde{B}_{6}=\big[H^{-1}(r)-1 \big]\, dy^{0}\wedge...\wedge dy^{6} \ , \qquad e^{2\phi}=H(r) 
   \ .
\end{align}
Here we find
\begin{align} \la{51}
    S_{\rm bulk}\Big|_{\rm NS5}= 0  
     \ ,& \qquad \ \ \ S_{\rm bndry}\Big|_{\rm NS5} 
    =-NT_{5}\int d^6y \ , \qquad 
        S_{\rm source}
   \Big|_{\rm NS5}=   -NT_{5}\int d^6y  
   \ ,\\ &  
\la{b10}     \qquad  S_{\rm tot}\Big|_{\rm NS5} 
   =-2NT_{5}\int d^6y \ .
\end{align}
Evaluating the bulk term  here and  in \rf{bb5}  we used  the explicit form of the  solution: 
note that  the  NS-NS   part of the bulk  action \rf{2}  or \rf{7}   automatically  vanishes  only for  solutions 
   without a  source term in the dilaton equation.

In the case of  D5-brane solution that has magnetic charge with respect to the RR 3-form $F_3$  we may again   introduce 
the dual electric   potential 
$\widetilde{C}_{6}$ ($d\widetilde{C}_{6}={}^{*}F_{3}$)     and consider  
\be
S_{source}=-N\mu_{5} \int d^{6}y \, e^{-\phi}\sqrt{G_{6}}+N\mu_{5}\int
\widetilde{C}_{6} \ , \ \ \ \qquad 
 \mu_5 = { 1 \ov  (2\pi)^5 \a'^3  g_s } \ . \la{b11} 
\ee
The   D5-solution  supported by the    corresponding  source at $x^a=0$ is \ci{Horowitz:1991cd}
\begin{align}
&d{s}^2=H^{-\frac{1}{2}}(r)\, \eta_{\m\n}dy^{\m}dy^{\n} +H^{\frac{1}{2}}(r)\,  dx^{a}dx^{a} \ , \qquad H(r)=1+\frac{Q}{r^2} \ ,
\quad \ Q=   { N \mu_5\k_{10}^2\ov \vol(S^3)}=\a' N g_s  \ , \no 
 \\
&\widetilde{C}_{6}=\big[H^{-1}(r)-1\big]\, dy^{0}\wedge...\wedge dy^{6} \ , \qquad e^{-2\phi}=H(r) \ .\la{b12}
\end{align}
The  resulting   contributions to the  total action \rf{b2} here   are 
\begin{align} \no 
    S_{\rm bulk}\Big|_{\rm D5} 
    =-\ha N\mu_{5}\int d^6y \ ,& \qquad     S_{\rm bndry}\Big|_{\rm D5}=-\ha N\mu_{5}\int d^6y \ , 
    \qquad S_{\rm source} \Big|_{\rm D5}= - N\mu_{5}\int d^6y \ ,  \\
    &\qquad S_{\rm tot}\Big|_{\rm D5}=-2 N\mu_{5}\int d^6y \ . \la{b15}
\end{align}
The  values of the total actions for NS5    \rf{b10}  and D5  \rf{b15}   cases 
  have the same  structure as for the D3-brane  solution  in \rf{39}
 and  also are consistent with the S-duality
relation between the two 5-branes.

Note  that the  bulk and boundary contributions  match only in sum:
 one can show  that the   S-duality transformation in the  formulation  using  the string frame  metric 
 leaves invariant only the  sum of the  bulk \rf{2}  and boundary \rf{99}  terms in the  type IIB action.



\small

\bibliographystyle{nb}
\bibliography{Boundary}

\end{document}

\bibliographystyle{JHEP}
\bibliography{Boundary.bib}

\end{document}